\documentclass[11pt,a4paper]{article}

\usepackage[utf8]{inputenc}
\usepackage[T1]{fontenc}
\usepackage{lmodern}
\usepackage{amsmath, amssymb}
\usepackage{graphicx}
\usepackage{hyperref}
\usepackage{longtable}
\usepackage{geometry}
\usepackage{url} 

\title{Integration of LSTM Networks in Random Forest Algorithms for Stock Market Trading Predictions}

\author{
  Juan C. King\\
  Centro de Investigación Operativa, Universidad Miguel Hernández, 03202 Elche, Spain\\
  \texttt{juan.carlos.king.perez@gmail.com}
  \and
  José M. Amigó\\
  Centro de Investigación Operativa, Universidad Miguel Hernández, 03202 Elche, Spain\\
  \texttt{jm.amigo@umh.es}
}

\date{2025}

\begin{document}

\maketitle

\abstract{The aim of this paper is the analysis and selection of stock trading systems that combine different models with data of different nature, such as financial and microeconomic information. Specifically, based on previous work by the authors and applying advanced techniques of Machine Learning and Deep Learning, our objective is to formulate trading algorithms for the stock market with empirically tested statistical advantages, thus improving results published in the literature. Our approach integrates Long Short-Term Memory (LSTM) networks with algorithms based on decision trees, such as Random Forest and Gradient Boosting. While the former analyze price patterns of financial assets, the latter are fed with economic data of companies. Numerical simulations of algorithmic trading with data from international companies and 10-weekday predictions confirm that an approach based on both fundamental and technical variables can outperform the usual approaches, which do not combine those two types of variables. In doing so, Random Forest turned out to be the best performer among the decision trees. We also discuss how the prediction performance of such a hybrid approach can be boosted by selecting the technical variables.}

\section{Introduction}

\label{s:intro}

\subsection{Objectives}

\label{s:objectives}

This paper is a follow-up to previous work by the authors on algorithmic trading in the cryptocurrency market using blockchain metrics and indicators \cite{king2024blockchain}. Here we extend that work to the stock market. In doing so, we draw on our experience with the cryptocurrency market, although the approach is necessarily different. More precisely, the main objective of this article is to implement an algorithmic intraweek trading system in the stock market by applying advanced artificial intelligence to fundamental and technical variables. An intraweek trading system consists of opening and closing trades over a period of several days or weeks \cite{chen2025investor}. 

In this context, the ability to accurately predict asset behavior is crucial
for investors. To meet this challenge, financial analysts and data
scientists have developed a variety of predictive models employing different
methods and datasets. In this article, we are going to analyze the following
three types of models.

\begin{description}
\item[(a)] \textit{Fundamental models} rely on key company
financial data, e.g., earnings per share (EPS), operating
cash flow, and market capitalization (\textit{fundamental variables}). Using advanced decision tree algorithms such as Random Forest and Gradient Boosting,  these models analyze a wide range of fundamental variables to predict the future behavior of assets. With a sufficiently large number of assets, these models offer a comprehensive view of the market and can identify significant patterns and trends influencing stock prices.

\item[(b)] \textit{Technical models} focus on stock price patterns and trading volumes (\textit{technical variables}). Using deep learning techniques such as
Long Short-Term Memory (LSTM) networks, these models incorporate a wide range of
technical indicators such as the Relative Strength Index (RSI), Moving
Average Convergence Divergence (MACD), and Bollinger bands. By processing
these technical signals, LSTM networks can capture complex patterns in stock prices and provide accurate predictions about their evolution.

\item[(c)] Models that combine fundamental and technical variables 
will be referred to as \textit{hybrid models}. In our approach, individual
LSTM nets are built for each asset, using both price data and technical
indicators. The output of these nets is then fed as an additional
variable into an algorithm such as Random Forest. This approach allows
to capture both long-term trends based on fundamentals and short-term
patterns based on technical analysis, thus providing a more comprehensive
view of the market and increasing prediction accuracy.
\end{description}

Whether based on fundamental variables, technical variables or a combination of both, the above models offer investors powerful tools to make
informed decisions and maximize their returns in a highly competitive and
volatile environment. In the present work, we study the integration of Random Forest and LSTM neural networks to obtain models with greater predictive capacity.

\subsection{Related work}

Predictive models in the stock market have evolved considerably thanks to
advances in data science and access to large datasets. In 2013, when the use
of Deep Learning algorithms and massive data processing and storage had not
yet exploded, algorithmic trading accounted for 52 \% of market order volume and 64 \% of nonmarketable limit order volume \cite{hendershott2013algorithmic}, i.e., algorithmic trading systems were monitoring market liquidity more actively than human traders. The first algorithmic trading systems were implemented with classic trading strategies, including MACD \cite{lei2020deep,mahajan2015optimization}, Bollinger bands \cite{prasetijo2017buy} and RSI. More recently, approaches to algorithmic trading have been 
proposed in which predictions are based on time series and indicators such as stochastic oscillators, moving averages, momentum, or Williams. Regarding fundamental variables, models based on EPS, company size, and more have been investigated in \cite{patell1976corporate,chan2021quantitative}. In these models, the objective is to replicate the manual trading strategies computationally. Nowadays, the increase in computational
power and the new algorithms that accompany it, allow thousands of variables
to be analyzed in real time.

In \cite{namdari2018integrating} a hybrid model was proposed in conjunction with a multi-layer perceptron. The data were stock
prices and financial ratios of technological companies listed on Nasdaq. In this case, the technical variables were discretized to combine them with the fundamental variables.

Combining predictive models is a field of extensive research in financial investment, as its use improves the possibilities and profitability of operations. Thus, in the work \cite{lazcano2023combined}, the dataset was analyzed using different cascade models, achieving better predictions. The proposed predictive model is a BiLSTM-GCN, a combination of directional LSTM networks and Graph Convolutional Networks (GCNs). In this case, no technical variables were used, only the price time series.

Similarly, in the work \cite{nobre2019combining} only a set of technical variables obtained exclusively from the prices was used. The number of variables was reduced with Principal Component Analysis (PCA), which facilitates data manipulation and reduces the computational resources required, thus improving the performance of the system. The system also improved the prediction accuracy metrics.

Unlike the previous two references, our predictive model uses both technical and fundamental data. This is why we need a different approach, which we call a hybrid model.

A different type of hybrid model is tested in \cite{lv2022stock}. In this case, algorithms are combined instead of fundamental and technical data. To be more precise, the authors combine several algorithms based on time series; if the first stage (consisting of Intrinsic Mode Functions (IMFs) with different scales and characteristics) surpasses a certain state, then an Autoregressive Moving Average (ARMA) is applied. 

Deep learning-based stock market prediction using price time series and LSTM neural networks is studied in \cite{pang2020innovative}. The algorithmic trading system proposed by the authors achieves a statistical advantage; the accuracy metric reported is 54.5 \%.

Finally, some models integrate predictive models with decision-making rules,
so they are two separate models rather than hybrid models. For example,
Yoon et al. (1994) proposed a model for the fundamental discrete variables and another for the search of patterns in time series.

A critical analysis of the existing literature reveals a persistent challenge: the effective integration of fundamentally distinct data modalities. While several works attempt hybrid approaches, they often rely on simplifications that compromise data integrity. Some models combine technical and fundamental variables but resort to discretization \cite{namdari2018integrating} or dimensionality reduction \cite{nobre2019combining}, inevitably losing granular information. Others focus on combining algorithms that process homogeneous data types, such as sequential time series \cite{lazcano2023combined, lv2022stock, pang2020innovative}, neglecting the integration of dynamic sequences with static multidimensional data. Crucially, no prior work addresses the native fusion of algorithms specifically designed for these intrinsically different data structures (sequential vs. static) without significant transformation or information loss. This represents a significant gap, as technical indicators (temporal dynamics) and fundamental factors (static context) capture complementary, yet structurally incompatible, aspects of market behavior.

To bridge this gap, our work introduces a novel hybrid architecture that achieves a breakthrough integration. Unlike previous approaches, our model preserves the native structure of each data type by seamlessly combining two specialized algorithms: one explicitly designed for processing sequential time-series data (technical indicators) and another optimized for handling static multidimensional variables (fundamental factors). This design allows for the first time a true synergistic analysis where temporal patterns and contextual fundamentals interact without reductive preprocessing, enabling a more comprehensive and accurate representation of market mechanisms. Our approach constitutes a significant advancement by directly addressing the core challenge of fusing heterogeneous data modalities at the algorithmic level.

\subsection{Contents}

This paper is organized as follows. In Section \ref{s:variables} we describe the variables used for the configuration of the models. They are of three types: fundamental variables (related to microeconomic data of companies), technical variables (related to indicators and oscillators), and the target variable (related to the predictions we want to make).

In Section \ref{s:methodology.3} we explain the methodology and quality metrics used in this work to measure the performance of the fundamental, technical and hybrid models briefly introduced in Section \ref{s:objectives} and detailed in Section \ref{s:MethodsPerformances.3}, after shortly explaining the preprocessing of the fundamental and technical variables in Section \ref{s:methods.1.0}.

Explicitly, we have chosen (i) decision trees (Random Forest and Gradient Boosting) and artificial neural networks for the implementation of fundamental models (Section \ref{s:methods.1.1}), and (ii) LSTM networks for the implementation of technical models (Section \ref{s:methods.3.2}). As for the hybrid models, in Section \ref{s:methods.3.3} we present the main novelty of this paper: a hybrid model whose inputs include outputs of technical models through the integration of predictions obtained from LSTM networks into the fundamental models. Last but not least, in Section \ref{s:methods.3.4} we discuss how to improve the performance of our hybrid model by selecting technical models with good prediction accuracy.

To complete the picture, in Section \ref{s:simulations} we simulate an investment strategy based on value (diversifying the entire capital across the 30 assets with the highest area under curve) over three weeks of the validation set. The results are compared with major indices (S\&P 500, NASDAQ, and Eurostoxx 50).

This paper ends with the main conclusions and a brief outlook in Section \ref{s:conclusion}.


\section{Variables}

\label{s:variables}

This section describes the fundamental and technical variables used in this article to implement predictive models. The prediction that these models are going to make is whether the price of a given asset will rise or fall in 10 business days (two weeks). To this end, in this section we also introduce a binary variable called target; its value at current time gives a projection of the price direction with the desired prediction horizon. 


\subsection{Fundamental variables}
\label{s:variables.1}

In this paper we use 32 fundamental variables. They represent a curated subset of hundreds of available metrics, rigorously selected based on their complementary information value, empirical predictive power, and coverage of essential financial dimensions (profitability, valuation, leverage, and market dynamics). This selection captures more than 90\% of the explanatory variance in equity returns according to fundamental factor studies \cite{damodaran2020equity}, maximizing informational content while maintaining algorithmic tractability.

The data consists of bimonthly microeconomic data from 482 companies, including tech giants like Apple, Microsoft and Nvidia; visit \cite{companylist} for a list of all these companies. This data was collected by a multinational financial data company that prefers to remain anonymous, and generously made available to us for the present work; similar data can be found in many platforms such as Alpha Vantage, Bloomberg or IEX Cloud, although they are all paid \cite{ng2022data}. Data collection for micro-economical data took place every 1st and 15th of each month in 2023, which amounts to a total of 482 datasets per company with its fundamental economical data. Price data, on the other hand, were collected from 1971 onwards, providing a long-term historical series calculation of indicators and technical models.


The 32 fundamental variables used in this study are designated below by a short name followed by a succinct description. They are organized by category.

\begin{enumerate}

\item \textbf{Profitability Ratios}
  \begin{enumerate}
  \item Div Yld Y0 Year Ended, Div Yld Y1 Current Year, Div Yld Y2, Div Yld Y3: Dividend Yield for the year ended (Y0) and the following three years (Y1, Y2, Y3). It represents the ratio of dividends paid per share to the share price.
  
  \item Margin Ebitda \% Y0 Year Ended, Margin Ebitda \% Y1 Current Year, Margin Ebitda \% Y2, Margin Ebitda \% Y3: EBITDA Margin for the year ended (Y0) and the following three years (Y1, Y2, Y3). It represents the ratio of EBITDA to total revenue, indicating profitability before interest, taxes, depreciation, and amortization \cite{okeke2020evaluating},
  \begin{equation*}
  \text{EBITDA Margin} = \frac{\text{EBITDA}}{\text{Total Revenue}} \times 100\%.
  \end{equation*}
  
  \item ROE Y1 Current Year: Return on Equity for the current year, representing the ratio of net income to shareholders' equity,
  \begin{equation*}
  \text{ROE} = \frac{\text{Net Income}}{\text{Shareholders' Equity}} \times 100\%.
  \end{equation*}
  
  \item Div Payout Y0 Year Ended, Div Payout Y1 Current Year, Div Payout Y2, Div Payout Y3: Dividend Payout Ratio for the year ended (Y0) and the following three years (Y1, Y2, Y3). It measures the percentage of earnings paid out to shareholders as dividends.
  
  \item EPS +1E 3Meses, EPS +1E Actual, var \% EPS +1E 3Meses: Next quarter estimated EPS, actual EPS, and percentage change in estimated EPS,
  \begin{equation*}
  \text{EPS} = \frac{\text{Net Income} - \text{Preferred Dividends}}{\text{Number of Common Shares Outstanding}}.
  \end{equation*}
  
  \item EBIT +1E 3Meses, EBIT +1E Actual, var \% EBIT +1E 3Meses: Next quarter estimated EBIT, actual EBIT, and percentage change in estimated EBIT.
  
  \item Sales +1E 3Months, Sales +1E Current, var \% Sales +1E 3Months: Next quarter estimated sales, actual sales, and percentage change in estimated sales.
  \end{enumerate}

\item \textbf{Valuation Ratios}
  \begin{enumerate}
  \item PER Y0 Year Ended, PER Y1 Current Year, PER Y2, PER Y3: Price-to-Earnings Ratio (PER) for the year ended (Y0) and the following three years (Y1, Y2, Y3). It measures the ratio of a company's share price to its earnings per share \cite{ghaeli2016price},
  \begin{equation*}
  \text{PER} = \frac{\text{Price per Share}}{\text{Earnings per share}}.
  \end{equation*}
  
  \item EV EBITDA Y0 Year Ended, EV EBITDA Y1 Current Year, EV EBITDA Y2, EV EBITDA Y3: Enterprise Value to Earnings Before Interest, Taxes, Depreciation, and Amortization (EV/EBITDA) multiple for the year ended (Y0) and the following three years (Y1, Y2, Y3). It indicates the valuation of a company relative to its operational cash flow \cite{zelmanovich2017basics},
  \begin{equation*}
  \text{EBITDA} = \text{Total Revenue} - \text{Variable Costs} - \text{Operating Expenses}.
  \end{equation*}
  
  \item Price to Book Value Y0 Year Ended, Price to Book Value Y1 Current Year, Price to Book Value Y2, Price to Book Value Y3: Price-to-Book Value Ratio for the year ended (Y0) and the following three years (Y1, Y2, Y3). It compares a company's market value to its book value \cite{budianto2023pengaruh},
  \begin{equation*}
  \text{P/B} = \frac{\text{Current Stock Price}}{\text{Book Value per Share}}.
  \end{equation*}
  
  \item Price CF Y0 Year Ended, Price CF Y1 Current Year, Price CF Y2, Price CF Y3: Price-to-Cash Flow Ratio for the year ended (Y0) and the following three years (Y1, Y2, Y3). It compares the market value of a company to its operating cash flow \cite{akhtar2015relationship},
  \begin{equation*}
  \text{P/CF} = \frac{\text{Current Stock Price}}{\text{Cash Flow per Share}}.
  \end{equation*}
  
  \item FCF/EV (\%) Y0 Year Ended, FCF/EV (\%) Y1 Current Year, FCF/EV (\%) Y2, FCF/EV (\%) Y3: Free Cash Flow to Enterprise Value Ratio for the year ended (Y0) and the following three years (Y1, Y2, Y3). It measures the percentage of free cash flow to the enterprise value \cite{toumeh2023surplus},
  \begin{equation*}
  \text{FCF/EV} = \frac{\text{Free Cash Flow}}{\text{Enterprise Value}}.
  \end{equation*}
  
  \item FCF YLD (\%) Y0 Year Ended, FCF YLD (\%) Y1 Current Year, FCF YLD (\%) Y2, FCF YLD (\%) Y3: Free Cash Flow Yield for the year ended (Y0) and the following three years (Y1, Y2, Y3). It represents the ratio of free cash flow per share to the share price \cite{purohit2025impact},
  \begin{equation*}
  \text{FCF Yield} = \frac{\text{Free Cash Flow}}{\text{Market Value of the Company}} \times 100\%.
  \end{equation*}
  
  \item PEG FY1, PEG FY2: Price/Earnings to Growth Ratio for the next year (FY1) and the following year (FY2). It relates the P/E ratio to the anticipated future earnings growth rate \cite{babii2023_nowcasting_pe},
  \begin{equation*}
  \text{PEG Ratio} = \frac{\text{Price/Earnings Ratio}}{\text{Annual EPS Growth Rate}}.
  \end{equation*}
  
  \item Objective Price 12 months, Potential Objective Price \%: 12-month price target and percentage potential for the price target. The "Potential Objective Price", also known as "Target Price" or "Target Price Potential", is an estimation of the future value of a financial asset, typically a stock. This calculation is based on various factors such as earnings forecasts, industry trends, market conditions, and other relevant information. Analysts and financial institutions often use different methodologies to derive target prices, including fundamental analysis, technical analysis, and valuation models \cite{damodaran2012investment},
  \begin{equation*}
  \text{Potential Objective Price} = \text{Current Price} \times \left(1 + \frac{\text{Growth Potential}}{100}\right).
  \end{equation*}
  
  \item Target Price 3Months, var \% PO 3Months: 3-month price target and percentage change in the price target.
  
  \item Long term growth \%: Long-term growth percentage. It is a financial metric used in financial analysis and company valuation to estimate the expected growth rate of a company's revenues, earnings, or other financial indicators in the future. This percentage represents the projected annual growth rate over an extended period, typically spanning several years. 
  \end{enumerate}

\item \textbf{Leverage Ratios}
  \begin{enumerate}
  \item Net Debt to EBITDA Ratio for the year ended (Y0) and the following three years (Y1, Y2, Y3). It measures a company's ability to pay off its debts using its earnings \cite{bouwens2019prevalence},
  \begin{equation*}
  \text{Net Debt to EBITDA Ratio} = \frac{\text{Net Debt}}{\text{EBITDA}}.
  \end{equation*}
  \end{enumerate}

\item \textbf{Market and Trading Data}
  \begin{enumerate}
  \item Market Value EUR millions: The total market value of a company's outstanding shares, expressed in millions of euros.
  
  \item Float Pct Total Outstdg: The percentage of total outstanding shares that are available for trading in the open market.
  
  \item Free-float EUR millions: The market capitalization of a company adjusted for the proportion of shares available for public trading, expressed in millions of euros.
  
  \item Recommendation and numerical recommendation for the stock: Recommendation and numerical recommendation for the stock. This indicator is obtained based on the analysis carried out by many traders and analysts manually.
  
  \item 12 months \%, YTD \%: Percentage change over the last 12 months and year-to-date, respectively.
  
  \item Last 52 Weeks Low Price and Last 52 Wks High Price: Lowest and highest prices over the last 52 weeks.
  
  \item \% From lows 1 year, \% from highs 1 year: Percentage change from 1-year lows and highs, respectively.
  
  \item 3y Price Volatility: Three-year price volatility.
  
  \item Issue Common Shares Outstdg, Average Daily Volume: Number of common shares outstanding and average daily trading volume, respectively.
  
  \item Volume/shares \%: Volume per share percentage.
  
  \item 3y BETA Rel to Loc Idx: Three-year Beta relative to the local index.
  
  \item \% Capital contracted daily: Percentage of capital traded daily.
  
  \item Diff \% Mean 200, diff \% Mean 50, diff \% Mean 25, Mean 50/200: Percentage difference from the 200 moving average daily timeframe.
  
  \item ECA Num EPS, ECA Num EBIT, EC Reco Total, EC Reco Up, EC Reco Down, EC Reco Unchng, \% mod recom Positivas/ Total, EC Reco Pos, EC Reco Neg, EC Reco Total.1, Positivas/ Total (\%): Various parameters related to earnings per share (EPS), Earnings Before Interest and Taxes (EBIT), and recommendations.
  \end{enumerate}

\end{enumerate}


\subsection{Technical variables}
\label{s:variables.2}

The technical variables used in this paper consist of the following 11 indicators and oscillators, grouped in three categories.

\begin{enumerate}

\item \textbf{Trend Indicators}
  \begin{enumerate}
  \item Simple Moving Average (SMA) \cite{huang2024prospective}. The formula is given by
  \begin{equation*}
  \text{SMA}_{n} = \frac{1}{n} \sum_{i=1}^{n} \text{Close}_{i}
  \end{equation*}
  where $n$ is the size of the time window (number of periods in the window), and $\text{Close}_{i}$ is the closing price at the $i$-th day of the window. $\text{SMA}_{n}$ was calculated for $n=20, 55$.
  
  \item Exponential Moving Average (EMA) \cite{formis2023linear}. The formula is given recurrently by
  \begin{equation*}
  \text{EMA}_{n} = \text{EMA}_{n-1} + \frac{2}{n + 1} \times (\text{Close}_{n} - \text{EMA}_{n-1})
  \end{equation*}
  
  where $n$ is the EMA span (lookback period), and $\text{Close}_{n}$ is the closing price at the last day of the period. $\text{EMA}_{n}$ was calculated for $n=20, 55, 200$. 
  
  \item Ichimoku Cloud \cite{lutey2022ichimoku} is defined by the formulas
  \begin{align*}
  \text{ITS}_{n} &= \frac{{\text{High} + \text{Low}}}{2} \quad (\text{over the last $n$ periods}) \\ 
  \text{IKS}_{m} &= \frac{{\text{High} + \text{Low}}}{2} \quad (\text{over the last $m$ periods}) \\
  \text{ISA}_{p} &= \frac{{\text{ITS} + \text{IKS}}}{2} \quad 
  \text{(over the last $p$ periods, plotted $\frac{p}{2}$)} \\
  \text{ISB}_{p} &= \frac{{\text{High} + \text{Low}}}{2} \quad (\text{over the last $p$ periods, plotted $\frac{p}{2}$}) \\
  \text{CS}_{m} &= \text{Closing Price} \quad (\text{plotted $m$ periods back})
  \end{align*}
  The "Tenkan Sen" line $\text{ITS}_{n}$ is called the conversion line; the "Kijun Sen" line $\text{IKS}_{m}$ is called the base line; the "Senkou Span A" line $\text{ISA}_{p}$ is called the leading span A; and the "Senkou Span B" line $\text{ISB}_{n}$ is called the leading span B; and the "Chikou Span" line $\text{CS}_{m}$ is called the lagging span. High and Low refer to the highest and lowest prices in the corresponding time window. We used the standard parametric values: $n = 9$, $m = 26$, and $p = 52$.
  
  \item Average Directional Index (ADX) \cite{salkar2021algorithmic} is calculated by the loop
  \[ \text{ADX}_{t} = \frac{\text{ADX}_{t-1} \times (n-1) + \text{DX}_{t}}{n} \] 
  where $n$ is the time window, $t=1,...,n$, 
  \[ \text{DX}_{t} = \frac{\left| \text{DI}^{+}_{t} - \text{DI}^{-}_{t}\right|}{\left| \text{DI}^{+}_{t} + \text{DI}^{-}_{t} \right|} \] 
  and
  \[ \text{DI}^{+}_{t} = \text{High}_{t} - \text{High}_{t-1} \] 
  \[ \text{DI}^{-}_{t} = \text{Low}_{t-1} - \text{Low}_{t} \] 
  where $\text{High}_{t}$ (resp., $\text{Low}_{t}$) is the highest (resp., lowest) price at period $t$. As usual, we set $n=14$.
  \end{enumerate}
  
\item \textbf{Momentum Oscillators}
  \begin{enumerate}
  \item Relative Strength Index (RSI) \cite{wu2015technical} is defined as
  \begin{equation*}
  \text{RSI}_{n} = 100 - \frac{100}{1 + \text{RS}}
  \end{equation*}
  where
  \begin{equation*}
  \text{RS} = \frac{\text{Average Gain}_{\,n}}{\text{Average Loss}_{\,n}}
  \end{equation*}
  
  Here, $\text{Average Gain}_{\,n}$ is the average gain over last $n$ days, 
  and $\text{Average Loss}_{\,n}$ is the average loss over last $n$ days. 
  RSI was calculated for $n = 6, 12, 14, 24$.

  \item Moving Average Convergence Divergence (MACD) \cite{kang2023optimal} is defined by the formulas
  \begin{align*}
  \text{MACD}_{n,m} &= \text{EMA}_{n}(\text{Close}) - \text{EMA}_{m}(\text{Close}) \\
  \text{MACDh}_{n,m,p} &= \text{MACD}_{n,m} - \text{EMA}_{p}(\text{MACD}_{n,m}) \\
  \text{MACDs}_{n,m,p} &= \text{EMA}_{p}(\text{MACD}_{n,m})
  \end{align*}
  where $n$ is the size of the time window for the shorter EMA, $m$ is size of the time window for the longer EMA ($m>n$), $p$ is the number of periods of the EMA signal line, $\text{EMA}_{n}(\text{Close})$ is the exponential moving average of the closing prices in the window of size $n$, $\text{MACDh}_{n,m,p}$ is the MACD histogram, and $\text{MACDs}_{n,m,p}$ is the MACD signal line. We used the standard parametric values: $n= 12$, $m= 26$ and $p=9$.
  
  \item Williams \%R \cite{steele2015technical} is given by
  \begin{equation*}
  \text{Williams \%R}_{n} = \frac{{\text{High}_{n} - \text{Close}}}{{\text{High}_{n} - \text{Low}_{n}}} \times 100\%
  \end{equation*}
  where $\text{High}_{n}$ is the highest price over the last $n$ periods, $\text{Low}_{n}$ is the lowest price over the last $n$ periods, and $\text{Close}$ is the closing price (at the last period). We used the standard value $n=14$.
  
  \item Stochastic Oscillator (KDJ) \cite{wu2015technical} is an indicator that measures the current price of an asset in relation to its range over a time interval. It is defined as
  \begin{align*}
  \%\text{K} &= \frac{{\text{Close} - \text{Min Low}_{n}}}{{\text{Max High}_{n} - \text{Min Low}_{n}}} \times 100\\
  \%\text{D} &= \text{SMA}_{n}(\%\text{K})\\
  \text{J} &= 3 \times \%\text{K} - 2 \times \%\text{D}
  \end{align*}
  where $\text{Close}$ is the current closing price, $\text{Min Low}_{n}$ is the asset lowest price over the last $n$ periods, and $\text{Max High}_{n}$ is the highest price over the same $n$ periods.
  We used the standard parametric values: calculation time interval for \%\text{K} = 14, and
  \%\text{D} smoothing window size = 3.
  
  \item Squeeze Momentum (SQZ) \cite{alostad2017directional} is a volatility indicator defined as
  \begin{equation*}
  \text{SQZ}_{n,m,p,q} = \frac{\text{SMA}(\text{Close}_{n}) - \text{SMA}(\text{Close}_{m})}{\text{SMA}(\text{Close}_{p}) \times q}
  \end{equation*}
  where $n$ is the size of the time window for the shorter mean (usually ranging from 5 to 20), $m$ is the size of the time window for the longer mean (usually between 20 and 50), $p$ is the size of the time window for the comparison mean, $q$ is a multiplier (typically 2 or 3), and $\text{SMA}(\text{Close}_{n})$ is the simple average of the closing prices in the window of size $n$. 
  We use the following parametric values: $n = 20$, $m = 50$, $p = 200$, and $q = 2$.
  \end{enumerate}
  
\item \textbf{Volatility Indicators}
  \begin{enumerate}
  \item Bollinger Bands \cite{su2023analysis} are envelopes plotted at a standard deviation level above and below a simple moving average of the price. They are defined by the formulas
  \begin{align*}
  \text{Lower band:} \;\; \text{BBL}_{n,k} &= \text{SMA}_{n}(\text{Close}) - k \times \text{StdDev}_{n}(\text{Close}) \\
  \text{Middle band:} \; \text{BBM}_{n,k} &= \text{SMA}_{n}(\text{Close}) \\
  \text{Upper band:} \;\; \text{BBU}_{n,k} &= \text{SMA}_{n}(\text{Close}) + k \times \text{StdDev}_{n}(\text{Close}) \\
  \text{Band width:} \;\; \text{BBB}_{n,k} &= \frac{\text{BBU}_{n,k} - \text{BBL}_{n,k}}{\text{BBM}_{n,k}} \\
  \text{Band percent:} \;\; \text{BBP}_{n,k} &= \frac{\text{Close} - \text{BBL}_{n,k}}{\text{BBU}_{n,k} - \text{BBL}_{n,k}}
  \end{align*}
  where $n$ is the size of the time window, $k$ is the number of standard deviations of the data in the window ($\text{StdDev}$), $\text{Close}$ is the closing price, $\text{StdDev}_{n}(\text{Close})$ is the standard deviation of the closing prices over the window, and $\text{SMA}_{n}(\text{Close})$ is the simple moving average of the closing prices over the window. 
  The parameter $n$ determines the sensitivity of the bands to changes in prices. The parameter $k$ determines the distance between the bands and the moving average. We used the standard parametric values: $n=20$ and $k=2$. 
  
  \item Average True Range (ATR) \cite{panapongpakorn2019short} is a price volatility indicator defined as
  \begin{equation*}
  \text{ATR}_{n} = \frac{1}{n} \sum_{i=1}^{n} \max(\text{High}_{i} - \text{Low}_{i},|\text{High}_{i} - \text{Close}_{i-1}|, |\text{Low}_{i} - \text{Close}_{i-1}|)
  \end{equation*}
  where $n$ is the size of the time window, $\text{High}_{i}$ is the highest price in the periods $1,2,...,i$, $\text{Low}_{i}$ is the lowest price in the periods $1,2,...,i$, and $\text{Close}_{i-1}$ is the closing price at period $i-1$. We used the standard value $n=14$.
  \end{enumerate}

\end{enumerate}

The selection of these 11 technical indicators was based on their widespread adoption and complementary functions:
\begin{itemize}
\item \textbf{Trend Indicators} (SMA, EMA, Ichimoku, ADX) identify market direction;
\item \textbf{Momentum Oscillators} (RSI, MACD, Williams \%R, KDJ, SQZ) measure price velocity and reversal points;
\item \textbf{Volatility Indicators} quantify price fluctuation intensity, specifically, Bollinger Bands measure volatility through band width (BBB) and band percent (BBP), while ATR measures average trading range regardless of direction.
\end{itemize}


The above 11 technical indicators constitute a validated core set selected from thousands of available tools based on their complementary signal properties, computational efficiency, and empirical effectiveness. As demonstrated in recent literature \cite{aronson2019evidence}, this combination provides optimal coverage of trend, momentum and volatility dimensions while capturing more than 85\% of the technical signal information value in equity markets.

All indicators use standard parameter settings as established in their original literature and industry practice, ensuring reproducibility and comparability with established financial research \cite{aronson2019evidence}.


\subsection{Target}
\label{s:variables.3}

Our target variable is the direction of stock prices in 10 business days. We will proceed formally and define a binary variable $\text{Target}(n)$.

Let $\text{P}(1), \text{P}(2), ..., \text{P}(T)$ be a time series of closing stock prices to be used in a prediction algorithm. To define the target variable at time $n$ (time at which the prediction is made) with prediction horizon $h$ (number of prediction periods, i.e., business days to make the prediction), such that $n+h \leq T$, we need first the so-called \textit{relative direction} $\text{Direct}(n+h)$, which is defined as
\begin{equation}
\text{Direct}(n+h) = \frac{\text{Max}_{P} - \text{P}(n)}{2} + \frac{\text{Min}_{P} - \text{P}(n)}{2} + \frac{\text{P}(n+h) - \text{P}(n)}{2}
\label{target}
\end{equation}
where $\text{P}(n)$ is the closing price at time $n$, $\text{Max}_{P}$ is the maximum closing price in the periods $n+1,..., n+ h$, and $\text{Min}_{P}$ is the minimum closing price in the same $n + h$ periods.  

In all tests performed in the present study, the default value of the prediction horizon is $h=10$. In view of Equation (\ref{target}), $\text{Direct}(n+10)$ assigns equal importance to the fluctuations between the initial closing price $\text{P}(n)$ and the maximum, the minimum, and the closing price of 10 days. This way, in the case of automated trading systems, it is possible to take profits before the close at 10 days if the price reaches a certain threshold or a stop loss is activated.

The Target at time $n$ for 10-day predictions, $\text{Target}(n)$, is then defined as
\begin{equation}
\text{Target}(n) = \begin{cases} 
1 & \text{if } \text{Direct}(n+10) - \text{P}(n) > 0 \\
0 & \text{if } \text{Direct}(n+10) - \text{P}(n) \leq 0 
\end{cases}
\label{target2}
\end{equation}
The discrete 0-1 variability of $\text{Target}(n)$ could have been smoothed with percentage increases, but for our purposes it suffices to know whether the price of the asset is going to rise ($\text{Target}(n)=1$) or fall ($\text{Target}(n)=0$). 


\section{Methodology}
\label{s:methodology.3}

Our study of algorithmic trading in the stock market started with the processing of historical microeconomic and stock price data of the 482 companies listed in \cite{companylist} to obtain the fundamental and technical variables presented in Sections \ref{s:variables.1} and \ref{s:variables.2}, respectively, for each company. With these datasets, we compared the 10-day stock price direction, measured by the Target variable (\ref{target2}) and predicted by the following models, which will be discussed in detail in Section \ref{s:MethodsPerformances.3}.

\begin{description}
\item[(M1)] \textit{Fundamental models}, which are machine learning algorithms based on decision trees and fed only with fundamental variables. 

\item[(M2)] \textit{Technical models} (one per asset), which are LSTM networks trained only with technical variables. 

\item[(M3)] \textit{Hybrid models}, which are models built on both fundamental and technical models.
\end{description}

Thanks to the use of a binary classifier (the Target) in the three cases, we could evaluate the performance of the above models with the same metrics, namely:

\begin{description}

\item[(i)] Area under the ROC curve (AUC).
\item[(ii)] Accuracy (ACC).
\item[(iii)] Recall (sensitivity).
\item[(iv)] Specificity.
\item[(v)] Precision.
\item[(vi)] F1-Score.
\item[(vii)] Type I Error (False Positive Rate).
\item[(viii)] Type II Error (False Negative Rate).

\end{description}

Furthermore, to detect overfitting, we resorted to the metrics.
\begin{eqnarray}
\text{Diff\ AUC} &=& \text{Train\ AUC} - \text{Test\ AUC} \label{AUCDIFF} \\
\text{Diif\ ACC} &=& \text{Train\ ACC} - \text{Test\ ACC} 
\end{eqnarray}
where Train AUC/ACC (resp., Test AUC/ACC) refer to results obtained with training (resp., testing) data.

To measure the prediction performance of models M1-M3 with the train and test datasets, we will use in Section \ref{s:MethodsPerformances.3} the Train and Test AUC as the main metrics. The reason for this choice is that, since the Target is a discrete variable, the AUC provides a view of the prediction quality of the models at a statistical level.


\section{Models and performances}
\label{s:MethodsPerformances.3} 

In this section, we describe the three models considered in this paper: the fundamental, technical and hybrid models. To be more specific, the hybrid model combines the best fundamental model (according to our performance results) and certain outputs of the technical models. Lastly, and most importantly, we will also describe how to improve the performance of the hybrid model by selecting the technical models.

Due to the limited availability of fundamental data and the architecture of our hybrid model, which depends on the output of previously trained LSTM models, the effective training window of the hybrid and fundamental models was restricted to 2023. The variables derived from technical models are obtained by training them from the initial listing date of each asset. For some assets, this corresponds to as early as 1971. The predictions generated for the year 2023 are subsequently incorporated into the fundamental or hybrid model.
As allocating a separate validation set within this limited range would have further reduced the training data, we opted to use all available data for training and testing. This trade-off was necessary to preserve the robustness of the hybrid model under data constraints. For the LSTM models, a chronological train-test split was implemented to prevent look-ahead bias. Approximately 60\% of each time series was used for training, 30\% for testing, and 10\% for validation. In cases where this condition could not be met, the corresponding records were excluded to ensure the integrity of the training process and avoid data leakage.

\subsection{Data pre-processing}
\label{s:methods.1.0}

This preprocessing approach was chosen as a compromise between data integrity and model coverage. The goal was to maximize the inclusion of informative records while maintaining consistency across the dataset. By eliminating unrecoverable data and applying simple imputation strategies where appropriate, we ensured that the dataset remained suitable for model training without introducing excessive noise. These choices reflect the practical constraints of working with real-world financial data, where missing information is frequent and must be addressed systematically to enable robust predictive modeling.

\subsubsection{Fundamental variables}

The preprocessing of the fundamental variables involved the following steps:

\begin{itemize}
    \item Due to the lack of sufficient data, the following columns were entirely removed:
    \begin{itemize}
        \item \texttt{capitalization\_millions}
        \item \texttt{float\_pct\_total\_outstdg}
        \item \texttt{free\_float\_eur\_millions}
    \end{itemize}
    
    \item For records with missing values:
    \begin{itemize}
        \item In the column representing the numerical analyst recommendation, missing values were replaced with a value of \textbf{1}, as it represents a neutral or average recommendation.
        \item In all other cases, missing values were replaced with \textbf{0}. This imputation was applied to variables such as \texttt{EBITDA}, \texttt{P/E ratio}, and \texttt{price-to-book ratio}.
    \end{itemize}
\end{itemize}

\subsubsection{Technical variables}

For technical variables related to price data, entries with missing values or non-existent records on individual days were removed from the dataset. Additionally, during the calculation of technical indicators and oscillators, these days were excluded from the computation. In cases where moving averages resulted in null values (typically at the beginning of the series), those records were also removed to ensure consistency in the input features.

\subsection{Fundamental models}
\label{s:methods.1.1}

For the implementation of predictive models using fundamental data with a 10-day prediction horizon, we employed three advanced machine learning algorithms: Random Forest \cite{breitung2023automated}, Gradient Boosting \cite{natekin2013gradient}, and a Neural Network model \cite{Goodfellow}. The models were implemented in Python 3.8 using Scikit-learn for tree-based models and Keras with TensorFlow backend for the neural network; hyperparameters were optimized using Random Search \cite{wang2025gaussian}. The best configurations obtained are described below.

\begin{itemize}

    \item \textbf{Random Forest (RF)}: Implemented with Scikit-learn’s \texttt{RandomForestClassifier}, the best model was obtained with the following parameters: \texttt{bootstrap=True}, \texttt{max\_depth=5}, \texttt{max\_samples=0.4},\texttt{max\_features="sqrt"}, \texttt{min\_samples\_leaf=13}, \texttt{min\_} \texttt{samples\_split=20}, \texttt{n\_estimators=170}, and \texttt{class\_weight="balanced\_subsample"}. This configuration achieved a Test AUC of 0.563 and a Test ACC of 0.543. 

    \item \textbf{Gradient Boosting (GB)}: The model was implemented using Scikit-learn’s \texttt{Gradient-} \texttt{BoostingClassifier}, which is a standard implementation of gradient boosting with decision trees. We did not use optimized variants such as LightGBM or XGBoost. The optimal configuration was: \texttt{learning\_rate=0.001}, \texttt{sub\_sample=0.4}, \texttt{max\_depth=3}, \texttt{max\_features="sqrt"}, \texttt{min\_samples\_leaf=8}, \texttt{min\_samples\_split=12}, and \texttt{n\_esti-} \texttt{mators=130}. This model achieved a Test AUC of 0.559 and a Test ACC of 0.560.

    \item \textbf{Neural Network (NN)}: Built using Keras and TensorFlow, the model consists of an input layer with one neuron per input feature and ReLU activation, followed by two hidden layers. The first hidden layer contains $n^2$ neurons (with $n$ being the number of input features), the second contains 256 neurons; both use ReLU activations and are regularized with Dropout (dropout rate = 0.3). The output layer has a single neuron with a sigmoid activation function for binary classification. The model was trained using the Adam optimizer with a learning rate of 0.01 and binary cross-entropy loss. The best configuration resulted in a Test AUC of 0.544 and a Test ACC of 0.534.

\end{itemize}

In view of the above Test AUC results, the a priori best model is Random Forest with the configuration given in Point 1. Therefore, this model will be the only fundamental model considered henceforth, also referred to as the \textit{fundamental best model}.

\begin{table}[htbp]
  \centering
    \begin{tabular}{lccc}

          & \textbf{Train} & \textbf{Test} & \textbf{Validation} \\

     \textbf{AUC}& 0.690 & 0.563 & 0.575\\
    \textbf{ACC (Accuracy)} & 0.633 & 0.543 & 0.560 \\
    \textbf{Recall (Sensitivity)} & 0.591 & 0.620 & 0.530 \\
    \textbf{Specificity}  & 0.673 & 0.483 & 0.588 \\
    \textbf{Precision}    & 0.634 & 0.482 & 0.552 \\
    \textbf{F1-Score}     & 0.612 & 0.543 & 0.541 \\
    \textbf{Type I Error} & 0.327 & 0.517 & 0.412 \\
    \textbf{Type II Error} & 0.409 & 0.380 & 0.470 \\
    \end{tabular}%
    \caption{Training, testing and validation metrics of the fundamental best model (Random Forest).}
  \label{tab:metrics.fundamental.overfitting}%
\end{table}%

Despite challenges in negative class identification, the model maintains reasonably consistent performance in key operational metrics, with only moderate degradation from training to testing. 
Most notably, the model maintains a stable F1-score (0.543 on the test set, 0.541 on the validation set), indicating a balanced precision--recall trade-off, while achieving an accuracy of 54.3\% on the test set and 56.0\% on the validation set. This combination of 62.0\% recall (effective opportunity capture) and 48.2\% precision (moderate signal quality) provides a practical foundation for upside-focused strategies, in which missing potential gains is costlier than incurring false positives.

While the observed specificity degradation (67.3\% $\rightarrow$ 48.3\% test, 58.8\% validation) and elevated Type I error (51.7\% test) suggest overfitting in negative class identification, these metrics must be interpreted within the context of bullish market conditions:
\begin{itemize}
  \item \textbf{Class imbalance naturally penalizes specificity}. The scarcity of negative cases (losses) limits the model's ability to generalize on their patterns.
  \item \textbf{Higher false positives are operationally tolerable} if the core objective remains maximizing upside participation (as evidenced by recall stability: 62.0\% on the test set, 53.0\% on the validation set)
\end{itemize}

From a generalization perspective, the drop in accuracy from train (63.3\%) to test (54.3\%) and validation (56.0\%) confirms a moderate overfitting tendency, primarily driven by weaker negative class detection. However, the closeness between test and validation metrics across recall, precision, and F1-score suggests that the model's behavior is stable when exposed to unseen data, supporting its operational reliability for real-time deployment in similar market regimes.

\begin{figure}[htbp]
  \centering
  \includegraphics[width=0.6\textwidth]{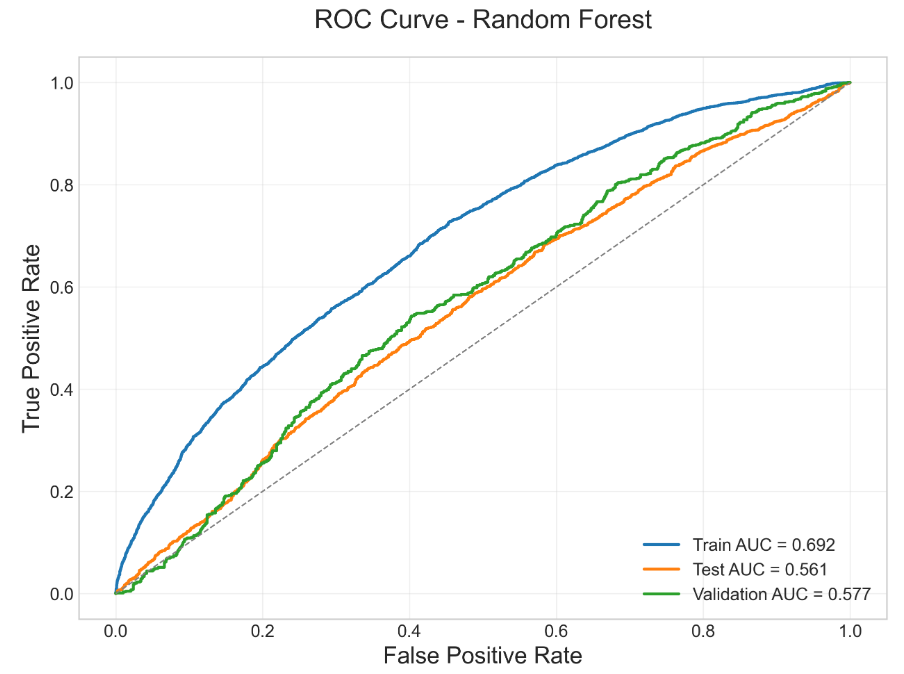}
  \caption{Train and Test ROC curves of the fundamental best model (Random Forest).}
  \label{fig:fundamental.auc}
\end{figure}

To complete the performance assessment, Table \ref{tab:metrics.fundamental.overfitting} "presents comprehensive performance metrics for the optimal fundamental model across training and testing datasets for 10-day-ahead forecasts." The corresponding ROC curves are shown in Figure \ref{fig:fundamental.auc}.

To further assess the model’s performance under a more robust validation framework, a rolling-window time series cross-validation was implemented, using 70\% of the data for training and 30\% for testing in each split. The area under the ROC curve (AUC) was computed for each of the five folds, yielding values of 0.5769, 0.5352, 0.5238, 0.5265, and 0.5962, with a mean of 0.5517 and a standard deviation of 0.0293. These results indicate a moderate yet consistent predictive performance across different temporal segments, supporting the stability of the model’s behavior in cross-validation.

\begin{figure}[htbp]
  \centering
  \includegraphics[width=0.6\textwidth]{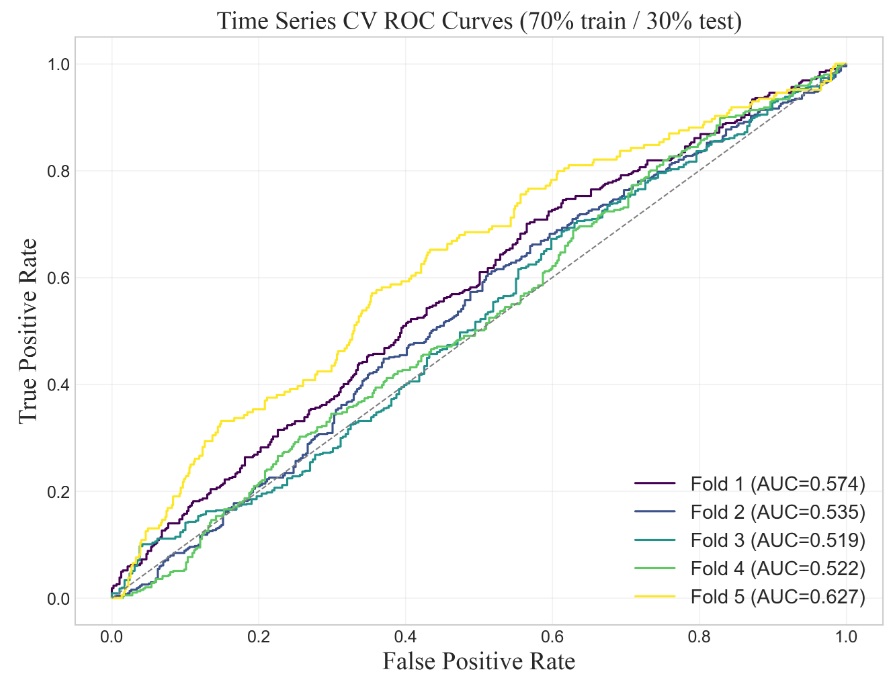}
  \caption{Train and Test ROC curves in cross validation (Random Forest).}
  \label{fig:crossVal.auc}
\end{figure}

The corresponding ROC curves for each fold are shown in Figure \ref{fig:crossVal.auc}.

\subsection{Technical models}

\label{s:methods.3.2}

In the case of technical variables (Section \ref{s:variables.2}), we use LSTM neural networks \cite{sherstinsky2020fundamentals} to forecast closing stock prices over a 10-day horizon. Since the time series of different assets have different patterns, the configuration of the network depends on the asset. Due to the complexity of LSTM networks, their adjustment and optimization was performed with the help of a search algorithm on a range of characteristics.

In our paper on blockchain indicators \cite{king2024blockchain}, we show that better results are achieved by converting the indicators into percentages with respect to the moving averages. Therefore, here we repeat the same procedure regarding stock prices with some indicators and oscillators. In doing so, we maintain the closing price data.

Furthermore, for each asset, we run a Greedy Search optimization \cite{garcia2025greedy} algorithm that, by using different configurations, obtained the best metrics based on the AUC. The feasible configurations were the following:

\begin{itemize}
\item Epochs: 20, 40, 60, 100.
\item Layers: 1, 4, 8.
\item Window: 10, 20, 30.
\end{itemize}

The best configuration on average was the one given by the triple 
\begin{equation}
\text{Epoch = 40},\; \text{Layers = 8},\; \text{Window = 30} 
\label{trinomial}
\end{equation}
This being the case, these are the corresponding settings of our technical models henceforth. Needless to say, other characteristics (notably the weights) depend on the given asset, so there is actually one technical model per asset.

All LSTM models were implemented in Python 3.8 using the Keras library with a TensorFlow backend. To measure the performance of technical models, we made predictions on train and test data. The aggregated results are given in Table \ref{tab:results.techinal.models}.

\begin{table}[htbp]
\centering
\begin{tabular}{lc}
\hline
\textbf{Metric} & \textbf{Av. ag. value} \\
\hline
Train AUC & 0.631 \\
Test AUC & 0.527 \\
Train ACC & 0.578 \\
Test ACC & 0.524 \\
Diff AUC & 0.104\\
Diff ACC & 0.053 \\
\hline
\end{tabular}
\caption{Average aggregated results for the technical models.}
\label{tab:results.techinal.models}
\end{table}

The mean Test AUC of 0.527 across 482 assets, while modest in absolute terms, achieves statistical significance and represents a consistent edge over random classification. The number of assets with $\text{Test\ AUC} \geq 0.50$ was 346 (71\%), the number of assets with $\text{Test\ AUC} \geq 0.60$ was 59 (12\%), and the number of assets with $\text{Test AUC} \geq 0.65$ was 18 (3\%). These metrics indicate that, at a general level, the models present a statistical advantage, although patterns with a high statistical advantage can only be found in a small group of assets.

The models were trained using the following hardware configuration: an HP Pavilion Gaming Laptop 15-ec0xxx equipped with an AMD Ryzen 5 3550H processor with Radeon Vega Mobile Graphics running at 2100 MHz, featuring 4 physical cores and 8 logical processors. The system ran Microsoft Windows 11 with 8 GB of RAM and 23 GB of virtual memory. The approximate training duration for each technical model ranged from 20 to 35 minutes, resulting in an overall experimentation process that lasted several weeks.


\subsection{A hybrid model}

\label{s:methods.3.3}

The hybrid model is a mixture of fundamental and technical models whose objective is to enhance the predictive capabilities of both types of models. To this end, we leverage predictions from already trained technical models as inputs to train a final model, the hybrid. Specifically, we feed three outputs from each technical model into the fundamental best model, additionally to the fundamental variables. 

\begin{figure}[htbp] 
    \includegraphics[width=0.95
    \textwidth]{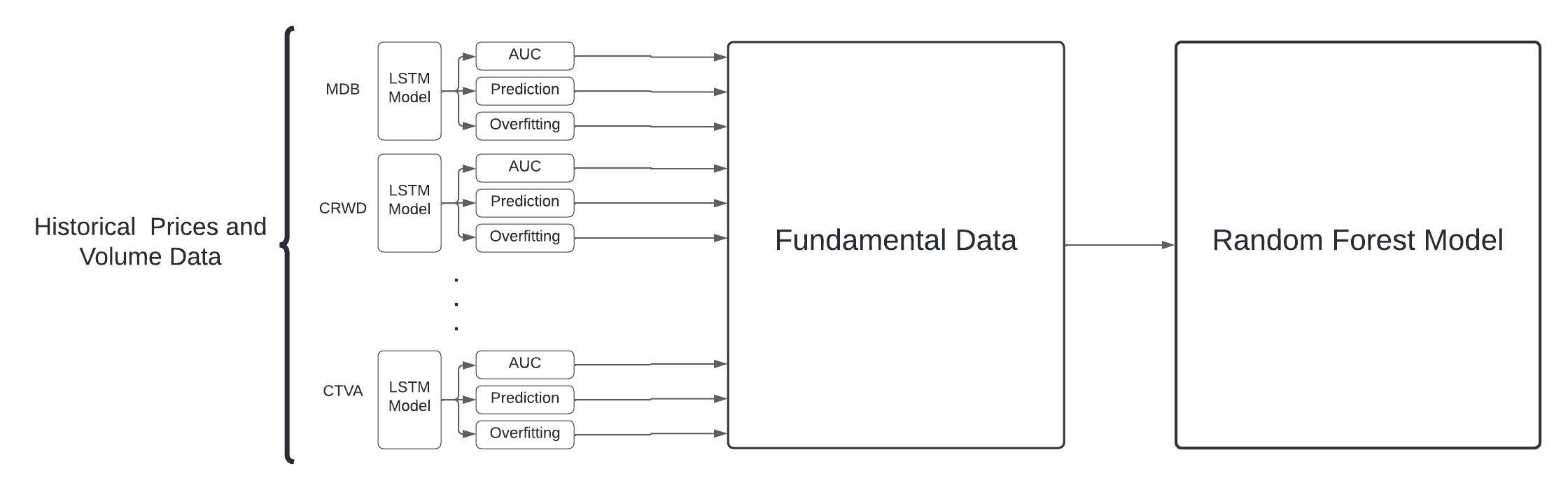}
    \caption{Scheme of the hybrid model.}
    \label{fig:hybrid_Model}
\end{figure}

Figure \ref{fig:hybrid_Model} illustrates the architecture of the hybrid model that we propose. Of course, the technical models are the 482 LSTM networks of Section \ref{s:methods.3.2}, trained with percentages and the closing stock prices of the 482 companies mentioned in Section \ref{s:variables.1} and listed in \cite{companylist}. The outputs of these technical models, fed into the Random Forest model along with the fundamental data of the same 482 companies, are the following.

\begin{description}

\item[(i)] Test AUC

\item [(ii)] The pondered (or weighted) probability, \text{Pond Prob}. This probability is calculated according to the formula
\begin{equation}
\text{Pond Prob} = \frac{{\text{Prob} - \text{Prob}_{\text{min}}}}{{\text{Prob}_{\text{max}} - \text{Prob}_{\text{min}}}}
\label{Pond Prob}
\end{equation}
where \text{Prob} is the probability that the price of the given asset will increase in 10 days, and $\text{Prob}_{\text{min}}$ (resp., $\text{Prob}_{\text{max}}$) is the minimum (resp., maximum) probability obtained in the test predictions.

\item [(iii)] Diff AUC (Equation (\ref{AUCDIFF})).

\end{description}

\begin{figure}[htbp]
  \centering
  \includegraphics[width=0.6\textwidth]{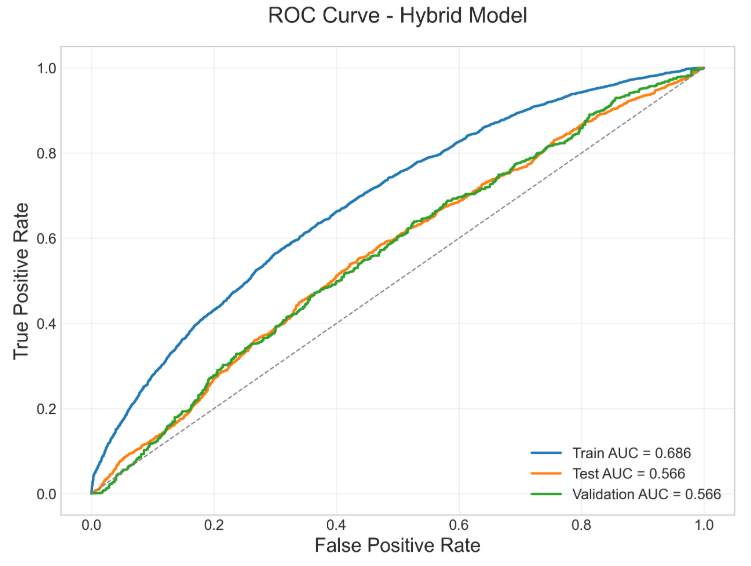}
  \caption{Train and Test ROC curves for the hybrid model.}
  \label{fig:hybrid.auc}
\end{figure}

The ROC curves of the hybrid model for training and testing data are shown in Figure \ref{fig:hybrid.auc}.

\begin{table}[htbp]
\centering
\begin{tabular}{lcccccc}

\textbf{Model}  & \textbf{Train AUC} & \textbf{Test AUC} & \textbf{Test AUC Sign.} & \textbf{95\% CI}\\

Fundamental & 0.690 & 0.563 & 0.563 & [0.548, 0.578]\\
Technical   & 0.631 & 0.527 & 0.493 & [-0.296, 1.282]\\
Hybrid      & 0.686 & 0.566 & 0.566 & [0.550, 0.581]\\

\end{tabular}
\caption{Comparative performance of the fundamental, technical and hybrid models (average values in the case of technical models), including Test AUC values, 95\% confidence intervals, and statistical significance results (Test AUC Sign.).}
\label{tab:analysis.models.fundamentals_hybrids}
\end{table}

\begin{figure}[htbp]
    \centering
    \includegraphics[width=0.9\textwidth]{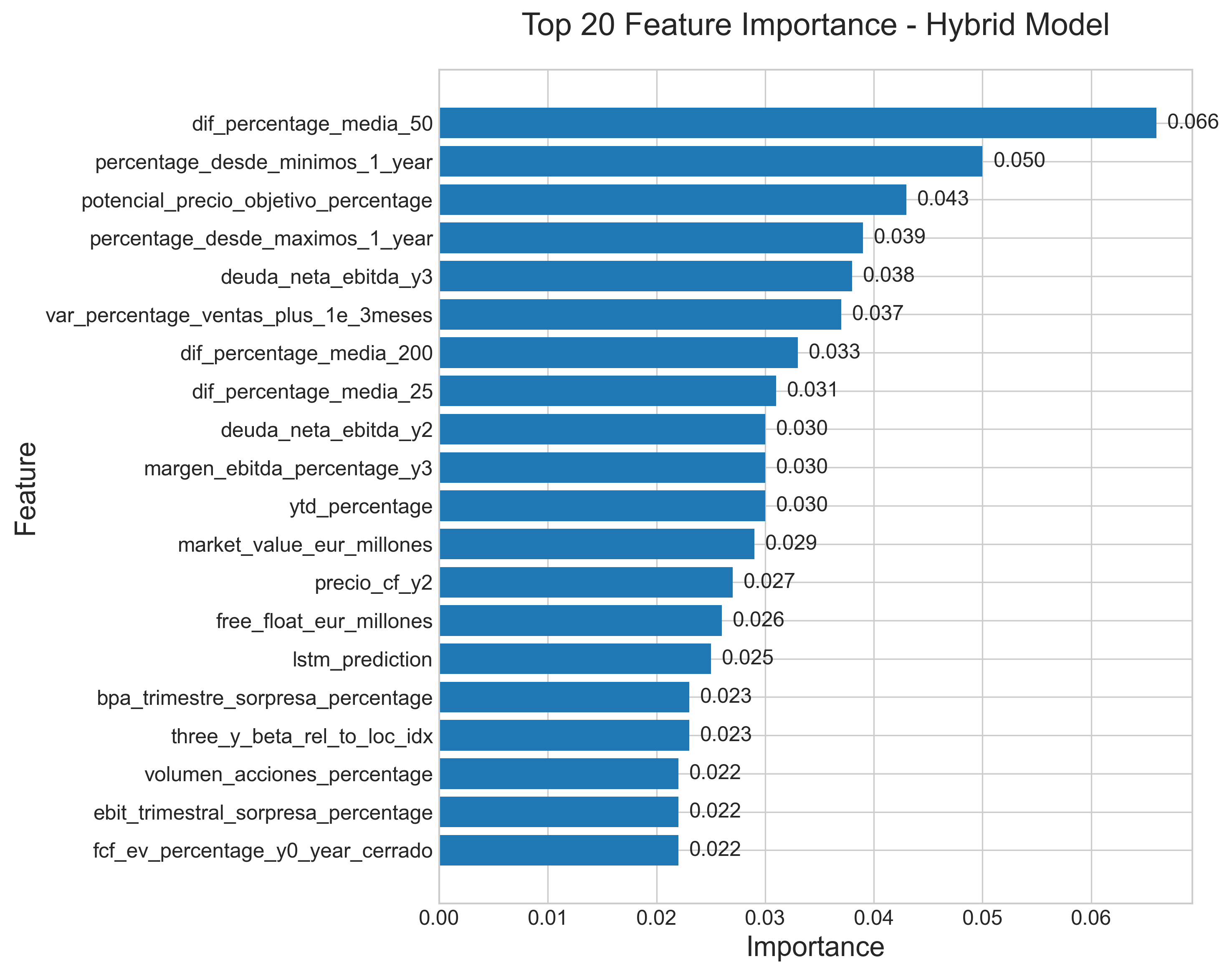}
    \caption{Top 20 feature importance ranking for the hybrid model. The variable "lstm\_prediction" is included at position 15 with an importance of 0.025.}
    \label{fig:feature_importance_top20}
\end{figure}

Table \ref{tab:analysis.models.fundamentals_hybrids} compares the performances of the best fundamental model (Table \ref{tab:metrics.fundamental.overfitting}), the technical models (Table \ref{tab:results.techinal.models}), and our hybrid model as measured by the Test AUC and Train AUC (average values in the case of the technical models). We see that the Test AUC of the hybrid model is 0.566, an increase of 0.003 with respect to the best fundamental model, and an increase of 0.071 with respect to the average Test AUC of the technical models. We conclude that the new data fed into the Random Forest in the hybrid model improves its predictive capability, albeit only slightly. As shown in Figure \ref{fig:feature_importance_top20}, the variable "LSTM prediction" ranks 15th in the variable importance ranking of the hybrid model, with an importance of 0.025076 ---a good value,. In the next section we explain how to boost the Test AUC metric of the hybrid model. Statistical significance tests were conducted to evaluate the predictive performance of the proposed models.

The Fundamental model achieved an AUC of 0.5628 with a 95\% confidence interval of [0.547, 0.578], indicating that its performance is statistically significantly better than random guessing. Similarly, the Hybrid model obtained an AUC of 0.566 with a 95\% confidence interval of [0.550, 0.581], also demonstrating statistically significant predictive capability. In contrast, the Technical model yielded an AUC of 0.493 with a wide 95\% confidence interval of [-0.296, 1.282], and therefore, no conclusion regarding its statistical significance can be drawn. These results confirm the effectiveness of the Fundamental and Hybrid models in outperforming random predictions, while the Technical model does not show evidence of predictive power beyond chance.

\subsection{An improved hybrid model}

\label{s:methods.3.4}

\begin{figure}[htbp]
  \centering
  \includegraphics[width=0.9\textwidth]{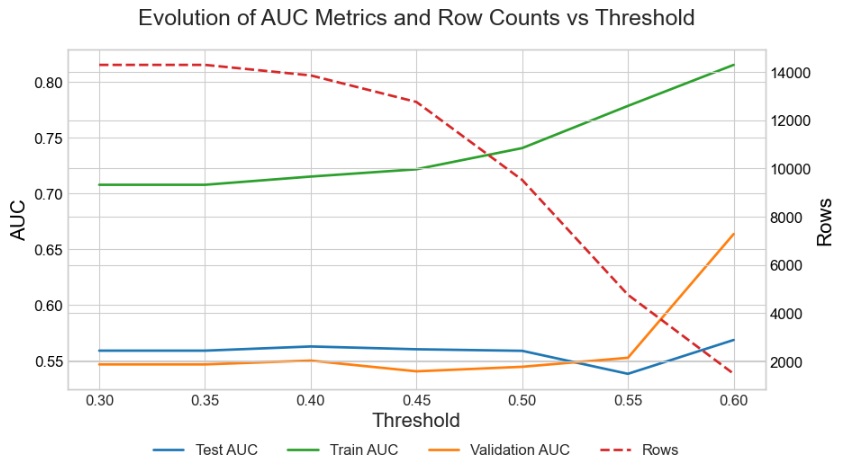}
  \caption{Evolution of the Train and Test AUC of the hybrid model vs the LSTM AUC threshold.}
  \label{fig:auc_evolution}
\end{figure}

Here we study the improvement of the hybrid model by selecting the technical models. For this purpose, we progressively filtered rows from the dataset based on LSTM-predicted values exceeding a threshold. This threshold corresponded to the AUC of the specific LSTM model and ranged from 0.3 to 0.6. All variables were retained; only rows not meeting the threshold were removed. For each filtered subset, we trained Random Forest models and measured their Train and Test AUC scores. Figure \ref{fig:auc_evolution} shows the evolution of these metrics versus the LSTM AUC threshold. 

We observe in Figure \ref{fig:auc_evolution} that, as the threshold increases, the number of rows of the dataset (discontinuous line) decreases while both the Train AUC (red continuous line) and the Test AUC (blue continuous line) increase. In fact, in the last iteration, where only 318 records with AUC greater than 0.6 remain, the Test AUC of the hybrid model is 0.73. This is a very high value for a trading system.

According to Figure \ref{fig:feature_importance_top10_filtered}, in the last filtered model, with an AUC greater than 0.6 in the LSTM prediction variable, this variable already occupies first place in the importance ranking of the model variables with an importance of 0.066, while it ranks 15th for the unfiltered hybrid model with an importance of 0.025 (Figure \ref{fig:feature_importance_top20}). 

\begin{figure}[ht!]
    \centering
    \includegraphics[width=0.9\textwidth]{feature_importance_top10_filtered.png}
    \caption{Top 10 feature importance ranking for the hybrid model filtered by LSTM AUC higher than 0.6. The variable "lstm\_prediction" occupies the first position with an importance of 0.066.}
    \label{fig:feature_importance_top10_filtered}
\end{figure}


\section{Simulation results}

\label{s:simulations}

In order to evaluate the effectiveness of the proposed system, a simulation was conducted on the validation set. The strategy was based on a value investing approach, selecting weekly the 30 assets with the highest AUC metric derived from the predictive model. Each asset was assigned the same initial capital, without leverage, and positions were held during the corresponding weekly period.

The returns obtained were compared with the performance of three major benchmark stock indices: the S\&P 500, Nasdaq Composite, and EuroStoxx 50. Over the three weeks analyzed, the proposed strategy demonstrated superior performance relative to these indices in terms of cumulative percentage return.

The weekly returns of the proposed algorithmic trading system were evaluated against three major market indices: the S\&P 500, Nasdaq Composite, and EuroStoxx 50, over a four-week period (week 0 to week 3). As shown in the figure, the first week (week 1) experienced a market downturn across all indices. Notably, the algorithmic system demonstrated greater resilience, exhibiting a smaller negative return compared to the indices, indicating its ability to mitigate losses during adverse market conditions. In the subsequent weeks (week 2 and week 3), the system outperformed all indices by achieving significantly higher positive returns, suggesting effective selection and allocation strategies. These results support the hypothesis that the investment approach, based on selecting the top 30 assets with the highest AUC and equally weighting positions, enhances risk-adjusted performance relative to broad market indices.

\begin{figure}[htbp]
  \centering
  \includegraphics[width=0.6\textwidth]{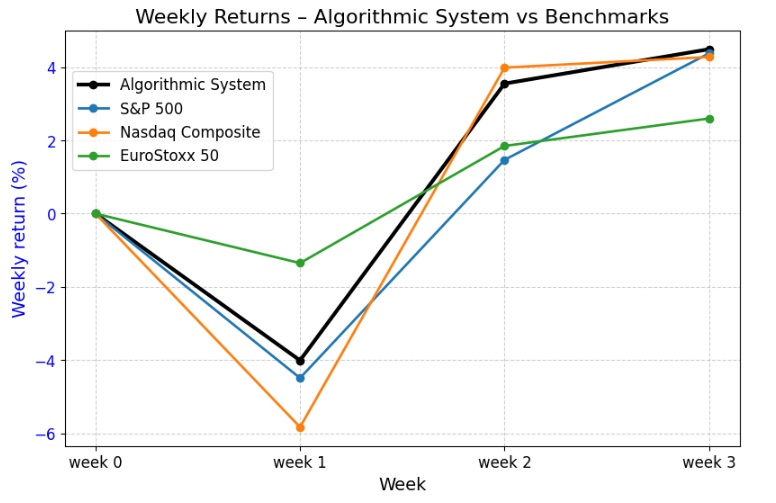}
  \caption{Comparison of algorithmic trading system with indices.}
  \label{fig:simulator}
\end{figure}

Figure \ref{fig:simulator} shows the weekly return comparison, where it can be observed that the system consistently outperforms the benchmark indices during the evaluated dates.

\section{Conclusion and outlook}

\label{s:conclusion}

The question studied in this paper is whether fundamental, technical or hybrid models provide a statistical advantage when trading stock market assets and which of them performs best. To answer this question, we converted it into a classification problem. The classes are defined by the binary variable $\text{Target}(n)$ at each prediction time $n$: $\text{Target}(n)=1$ means that the price of the asset will increase in 10 business days, whereas $\text{Target}(n)=0$ means the opposite.

The type of model refers to the data used. Thus, for the fundamental models we used microeconomic data of 482 companies in year 2023. These data provided good predictive capability. For the technical models, we used indicators and oscillators based on stock price and volume data from the same companies. We found that, in some cases, technical models also provided good predictive capabilities. As their name suggests, hybrid models are a combination of fundamental and technical models.

For fundamental models we considered three candidates: Random Forest, Gradient Boosting and Neural Network. The selection criterion consisted of predicting, with a 10-day objective, which companies were the best to invest in. The fundamental model that performed best (the "fundamental best model") was Random Forest, with the hyperparameters given in Section \ref{s:methods.1.1} and a Test AUC of 0.563.

The technical models were implemented with LSTM networks, one per asset. The task to measure their performance consisted of predicting, also with a 10-day objective, whether the price of the corresponding asset was going to rise or fall. We found that the models built using price percentages with respect to indicators provided more accurate predictions. In these technical models, a statistical significance test was performed on all aggregated models, and the result was negative, with an AUC of 0.493 and a wide 95\% confidence interval of [-0.296,1.282].  In models where the AUC is below 0.5 or the AUC difference is excessively high, there is a risk of overfitting. This may indicate that the integration of variables obtained from the LSTM networks does not lead to a substantial improvement of the first hybrid model (Section \ref{s:methods.3.2}) over the fundamental model.

As for hybrid models, the model implemented in this paper is made up of the fundamental best model (Random Forest), additionally fed with information output by the technical models (three inputs per technical model, i.e., per asset). As a result, the Test AUC of the Random Forest increased from 0.563 (with only fundamental data) to 0.566 (with additional technical data). This finding suggests that the technical features derived exclusively from the LSTMs are insufficient to enhance the model.

Test data were obtained using 20\%, and validation AUC data were obtained using 10\% of the dataset. The results were comparable between the fundamental and hybrid models, with the fundamental model achieving AUC values of 0.690 (train), 0.563 (test), and 0.577 (validation), and the hybrid model achieving 0.686, 0.566, and 0.566, respectively. These findings suggest that both approaches exhibit similar generalization capabilities, with no substantial improvement observed when integrating hybrid features over the purely fundamental approach. Nevertheless, it is worth noting that the variable \textit{lstm\_prediction} ranked 15\textsuperscript{th} in the feature importance analysis, indicating that it was a relatively important variable in the hybrid model.

This being the case, in Section \ref{s:methods.3.4} we demonstrated that the quality of the hybrid model (as measured by the AUC) can be further improved by reducing the number of records, in fact, by operating with LSTM networks that have a Test AUC greater than 0.6. In this situation, while the universe of assets decreases, the statistical advantage of the model increases, reaching an AUC of 0.73. In addition, the variable obtained through such LSTM networks was one of the 10 most important, according to Figure \ref{fig:feature_importance_top10_filtered}. This point is crucial because it indicates that, if the best technical models (LSTM networks) are part of the hybrid model, the statistical advantage increases considerably, and this is one of the most important objectives in algorithmic trading.

LSTM models are highly complex and computationally intensive models. Ideally, a hyperparameter optimization algorithm tailored to each asset would have been more appropriate, rather than applying a single generalized configuration. However, given the large number of assets (482), we were compelled to generalize and seek a configuration that was optimal overall. 
In this case, a pattern search algorithm is needed for each asset, and the optimization of the corresponding algorithm requires a different number of layers, epochs, window sizes and hyperparameter settings. Therefore, it is unfeasible in practice to generalize a solution for all assets. In our case, we focused on a limited number of possible solutions, certainly very small compared to the huge number of algorithm configurations.

Another important point is the quality of the data and its cost. In general, the provision of fundamental data sets is expensive and not within reach of small traders. This can mean a significant disadvantage for the latter compared to large corporations. And yet, we showed in the previous sections that it is relatively easy to obtain a statistical advantage with fundamental models when trading in the stock market.

We have also shown that technical data (basically, stock price and trading volume) can furnish predictive algorithms with sufficient statistical advantage. Since that type of data is available to any trader, so is profitability via technical models. The problem with technical models based on LSTM networks is that they predict a single asset and, therefore, the risk is very high. Indeed, an unforeseen event that causes the price to move in the opposite direction than that expected can lead to large capital losses. For this reason, it is advisable to use hybrid models that allow for the diversification of positions.

It is likely that the optimal architecture for each individual asset was not found, as different assets exhibit different patterns and may require customized configurations. Implementing an algorithm capable of identifying the best architecture per asset could have resulted in significantly improved performance—potentially achieving AUC scores above 0.6 for most assets. However, such an optimization process would require substantial computational infrastructure, which was beyond the scope and resources of this study. While filtering the asset universe to those with LSTM test AUC > 0.6 may improve predictive accuracy, it also significantly narrows the investable pool, potentially compromising diversification and increasing concentration risk. Furthermore, regulated investment frameworks—such as the UCITS Directive (2009/65/EC)—mandate that no more than 5 \% of a fund’s assets may be invested in transferable securities or money market instruments issued by the same body \cite{kasyanov2024evolution}. This regulatory constraint illustrates why a sufficiently broad universe of eligible assets is necessary to comply with concentration limits and ensure robust risk management. Therefore, it becomes essential to continue improving this contribution by developing techniques that increase the AUC across all LSTM models, in order to make such a system viable and implementable in real-world financial environments.

One potential direction for future research involves replacing the LSTM networks with Graph Neural Networks (GNNs), which may offer a more powerful representation of relationships among financial variables. GNNs have shown promising results in domains where the structure of data can be modeled as a graph, and this approach could help improve the model’s predictive performance (e.g., AUC). However, implementing GNNs comes with significantly higher computational costs and infrastructure requirements, as they require explicit graph construction, greater memory consumption, and more complex training pipelines. Despite these challenges, their ability to capture latent dependencies between assets or macroeconomic indicators could be valuable in real-world financial applications. 

While the results presented demonstrate promising improvements with the hybrid modeling approach, some limitations should be noted. First, the computational cost of training separate LSTM networks for each asset can be substantial, especially as the number of assets grows, which may limit scalability and real-time applicability in larger universes. Second, our study focuses primarily on a specific market (e.g., developed markets), and the generalizability of these findings to other contexts, such as emerging markets with different market dynamics, liquidity profiles, and data availability, remains to be validated. In particular, emerging markets often present additional challenges due to the scarcity of sufficient historical data, which makes it more difficult to build statistically robust models. Furthermore, creating LSTM models to cover all assets across emerging markets would require extremely high storage and processing capabilities. Moreover, LSTMs themselves struggle to be effectively trained on very long sequences, which further complicates their application in these markets. Despite these difficulties, emerging markets hold significant potential due to their higher volatility and possible opportunities for greater predictive gains. Importantly, the proposed hybrid system could be adapted and applied to other similar markets, such as Asian or South American markets, where similar conditions and challenges exist. Future work could explore strategies to reduce computational demands—such as transfer learning or model sharing—and extend the analysis to diverse markets to assess robustness and adaptability.

The results of the market application simulator for validation data simulating an Algorithmic Trading System indicate that the algorithmic system operates in line with the main market indices, but slightly outperforms them during the three-week test period. However, to draw robust conclusions about its long-term effectiveness, a comprehensive evaluation over multiple years and diverse market conditions would be necessary.

As a final message, the combination of fundamental and technical models into hybrid models with improved predictive capabilities is a topic worth investigating in algorithmic trading. Actually, in this paper we have proved the feasibility of that approach. In addition to new ways of defining hybrid models with ever better performances, future work also includes the study of more cognitive models taking into account macroeconomic data, responses to news or events, psychological supports and resistances, Fibonacci levels or, for that matter, any variable that could have a correlation with the price direction. Other related topics are the optimization of the LSTM nets, as well as the impact of taking different targets, time-frames and combinations of both.


\end{document}